\documentclass{aa}
\usepackage{psfig} 
\begin{document}

\title{XMM-Newton monitoring of X-ray variability in the quasar 
\object{PKS~0558--504}
\thanks{Based on observations with XMM-Newton, an ESA Science Mission 
    with instruments and contributions directly funded by ESA Member
    States and the USA (NASA).}}
  
\author{M. Gliozzi\inst{1}  
 \and W. Brinkmann\inst{1} 
 \and P.T. O'Brien\inst{2}
 \and J.N. Reeves\inst{2} 
 \and K.A. Pounds\inst{2} 
 \and M. Trifoglio\inst{3}
 \and F. Gianotti\inst{3}}
\offprints{mgliozzi@xray.mpe.mpg.de} 
\institute{
Max-Planck-Institut f\"ur extraterrestrische Physik,
         Postfach 1312, D-85741 Garching, Germany
\and X-ray Astronomy Group, Department of Physics and Astronomy,
University of Leicester, LE1 7RH, U.K.
\and Istituto TESRE, CNR, Via Gobetti 101, I-40129 Bologna, Italy}
\date{Received: ; accepted: }

\abstract{
We present the temporal analysis of X-ray observations of the radio-loud
Narrow-Line Seyfert 1 galaxy (NLS1) \object{PKS~0558--504} obtained during the
{\it XMM-Newton} Calibration and Performance Verification (Cal/PV) phase. 
The long term light curve is characterized by persistent variability with 
a clear tendency for the X-ray continuum to harden when the count rate 
increases. Another strong correlation on long time scales has been found 
between the variability in the hard band and the total flux. On shorter 
time scales the most relevant result is the presence of smooth modulations, 
with characteristic time of $\sim$ 2 hours observed in each individual 
observation. The short term spectral variability turns out to be rather 
complex but can be described by a well defined pattern in the hardness 
ratio--count rate plane.
}
\titlerunning{XMM-Newton observations of \object{PKS~0558--504}}
\authorrunning{M.~Gliozzi et al.}
\maketitle
\section{Introduction}
Active Galactic Nuclei (AGN) are variable in every observable wave band.
The X-ray flux exhibits variability on shorter time scales than any other
energy band, indicating that the emission occurs in the innermost regions
of the central engine. Therefore, a study of the X-ray variability is a
powerful tool providing upper limits on the sizes of the emitting
regions and the masses of central black holes and it allows one to probe the 
extreme physical processes operating in the inner parts of the accretion flow 
close to the event horizon. Although X-ray variability has been observed in AGN
for more than two decades, its origin is still poorly
understood. Narrow-Line Seyfert 1 galaxies (NLS1) often display rapid,
large amplitude X-ray variability as well as extreme long term changes
(Forster \& Halpern \cite{forst}, Boller \cite{boll1}, Brandt \cite{brand}), and therefore
they represent the ideal class for an X-ray temporal analysis.
\begin{table} 
\caption{EPIC PN individual observations.}
\begin{center}
\begin{tabular}{llll}
\hline
\hline
\noalign{\smallskip}
Observation date & Duration& Mode$^a$ & Filter$^b$\\
\noalign{\smallskip}       
\hline
\hline
\noalign{\smallskip}
\noalign{\smallskip}
00/2/7 UT11:15-13:19 & 7.2 ks & FF & M\\
\noalign{\smallskip}
\hline
\noalign{\smallskip}
00/2/7 UT14:28-16:02 & 5.4 ks & LW & M\\
\noalign{\smallskip}
\hline
\noalign{\smallskip}
00/2/10-11 UT23:26-3:23 & 14.0 ks & FF & T2\\
\noalign{\smallskip}
\hline
\noalign{\smallskip}  
00/3/2 UT18.15-21:42 & 12.4 ks & FF & M\\
\noalign{\smallskip}
\hline
\noalign{\smallskip}
00/5/24 UT17:21-21:07 & 13.6 ks & SW & M\\ 
\noalign{\smallskip}
\hline 
\end{tabular}
\end{center}
$^a$ Full Frame (FF), Large Window (LW), Small Window (SW).\\
$^b$ Medium (M), Thin 2 (T2).\\
\end{table}
\object{PKS~0558--504} ($z=0.137,~m_{\rm B}=14.97$) is one of the few radio-loud
NLS1. In X-rays it is characterized by
a steep spectrum [the photon index $\Gamma$ ranges between 2.2 and 3.1, 
obtained from different
instruments and energy bands (a detailed spectral analysis of {\it XMM-Newton}
observations has been performed in a separate paper, O'Brien \cite{obri})], 
high luminosity [$(2-5)\times
10^{45}{~\rm erg~s^{-1}}$, with $H_0=70 {~\rm km~s^{-1}~Mpc^{-1}},~
q_0=0.5$], and strong variability. A Ginga observation
(Remillard \cite{remi}) showed an increase of the X-ray flux by $67\%$
in 3 minutes, implying the presence of a relativistic flare. Recent 
ROSAT High Resolution Imager (HRI) observations (Gliozzi \cite{glioz})
confirmed the presence of strong and persistent
X-ray variability, which suggests the presence of a 
rotating black hole, and ruled out external contributions
to the high luminosity and variability from other nearby sources.

In this paper we report the results of four {\it XMM-Newton} observations
of \object{PKS~0558--504} taken with  the European Photon Imaging Camera (EPIC) PN,
which is the best suited instrument on board the ESA satellite
for timing analysis purposes, due to its high time resolution.
In Sect.2 we describe the observations and data reduction.
The long and short term X-ray variability are discussed in Sect.3 and Sect.4,
respectively. Sect.5 contains the main conclusions and a summary.

\section{Observations}
\object{PKS~0558--504} was observed with the EPIC PN camera
during orbits 30, 32, 42 and 84 (corresponding to February 7 and 10,
March 2 and May 24, respectively). The camera was operated in different 
observing modes and with various filters. 
In the Full Frame mode all the pixels of all CCDs are read 
out with time resolution of 70 ms and the full field of view 
($\sim$ 30\arcmin) is covered. In the Partial Window modes (Large Window and
Small Window) a two dimensional imaging readout over part of the CCDs array
is performed with increased time resolution (48 ms and 6 ms, respectively).
The primary reason for using filters is that the EPIC CCDs are not only
sensitive to X-ray photons, but also to IR, visible and UV light. Therefore
for objects with high optical to X-ray flux ratio, the X-ray signal might be
contaminated by those photons. This is not the case of \object{PKS~0558--504}, for which
different filters were used only for calibration purposes.
Table 1 provides a summary of the individual observations. 
The data reduction and analysis were performed using the last release
of the {\it XMM-Newton} SAS (Science Analysis Software) package. 
From the available 
observations, the last reprocessed data sets without any indication of timing 
problems  
and only single events (i.e. events where the charge released by a photon is
contained in a single CCD pixel) have been analyzed. The main reason for
using only singles is related to the energy dependence in the relative
fraction of ``multiple" events. To quantify their contribution,
a test with all valid events has been performed. 
The main effect is an increase of the count rate by $\sim10\%$ and variations
of the hardness ratio by $\sim20\%$, which would introduce additional 
uncertainties. Light curves were
obtained by extracting photons from circular region of radius of
$\sim$ 33\arcsec ~around the source center and subtracting the background 
taken from a source-free region (which amounted to $\sim 1\%$). After  
confirming the lack of significant
variability on time scales less than minutes, the photons were binned into
200 s intervals providing a good signal to noise ratio.
\object{PKS~0558--504}
was chosen as a CAL/PV target with the purpose of investigating the photon
pile-up in the EPIC cameras. With this aim we compared the spectra from the 
central pixels with those from the outer parts of the PSF and found no
significant indication for photon pile-up.

\section{Long term variability}
\begin{figure}
\psfig{figure=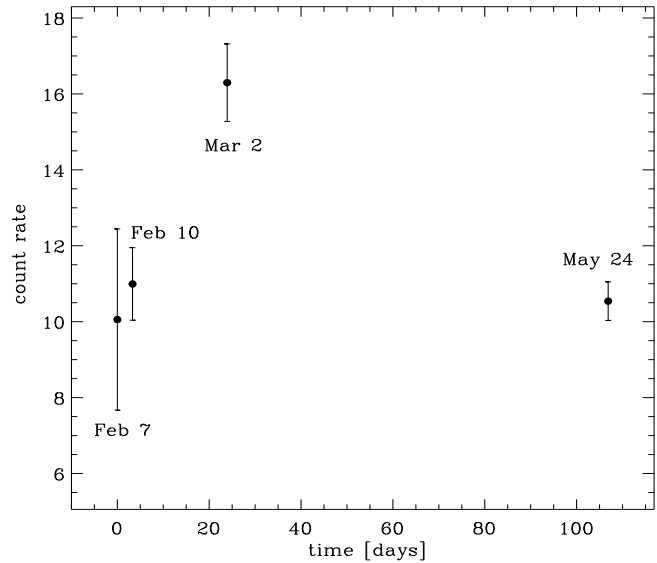,height=7.5cm,width=8.7cm,%
bbllx=40pt,bblly=120pt,bburx=550pt,bbury=600pt,clip=}
\caption{EPIC PN 0.2-10 keV light curve of \object{PKS~0558--504} for orbit 30, 32, 42
and 84. 
\label{figure:XMM57_f1.eps}}
\end{figure}
In  Fig.~\ref{figure:XMM57_f1.eps} we show the total light curve for \object{PKS~0558--504} 
from February 7 to May 24 2000. Each data point
is an average count rate over the individual observation in the range
0.2-10 keV. The indicated error of $1\sigma$, represents the dispersion 
around the mean and gives an idea of
the variability within each observation. The error bar is particularly large
for the first data point because we merged two exposures (with very
different mean count rates) taken during orbit 30 nearly one hour apart. 
The total light curve is characterized by a moderate variability, if compared
with the ROSAT light curve where the count rate varied by a factor 5 in less
than 3 days (Gliozzi \cite{glioz}). In the EPIC PN light curve the most 
prominent variation in the count rate is by a factor of 2
over a period of 20 days.

\begin{figure}
\psfig{figure=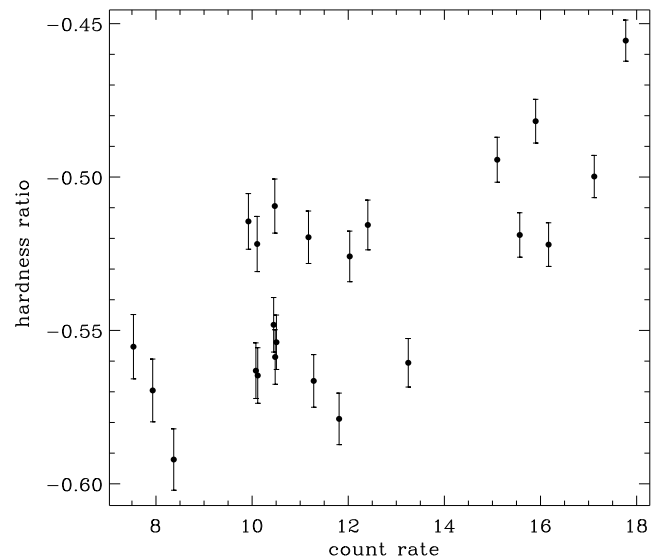,height=7.5cm,width=8.7cm,%
bbllx=10pt,bblly=120pt,bburx=550pt,bbury=600pt,clip=}
\caption{Hardness ratio (hard-soft)/(hard+soft) versus total count rate.
Every data point corresponds to 2000 s integration time. Soft band (0.2-1 keV),
hard band (1-10 keV).
\label{figure:XMM57_f2}}
\end{figure}

We sought evidence for spectral variability by computing the hardness ratios
as a function of the total 0.2-10 keV flux. As hardness ratio we define
the difference between the count rates in the $1 {~\rm keV}\leq {\rm E}
\leq 10{~\rm keV}$ band and those in the $0.2 {~\rm keV}\leq {\rm E}<
 1{~\rm keV}$ band, divided by the 0.2-10 keV count rate. These ranges where chosen
 to provide a good signal to noise in both the soft and the hard light curves.
The hardness ratios versus the total count rate with data from each observation
binned in 2000 s intervals is shown in Fig.~\ref{figure:XMM57_f2}. 
The presence of a positive
correlation was quantitatively tested by performing a linear least square fit,
which confirmed the result at 5$\sigma$ confidence level. On long time scales
the spectral variability is correlated with flux variations such that the
spectrum becomes harder when the count rate increases.

\begin{figure}
\psfig{figure=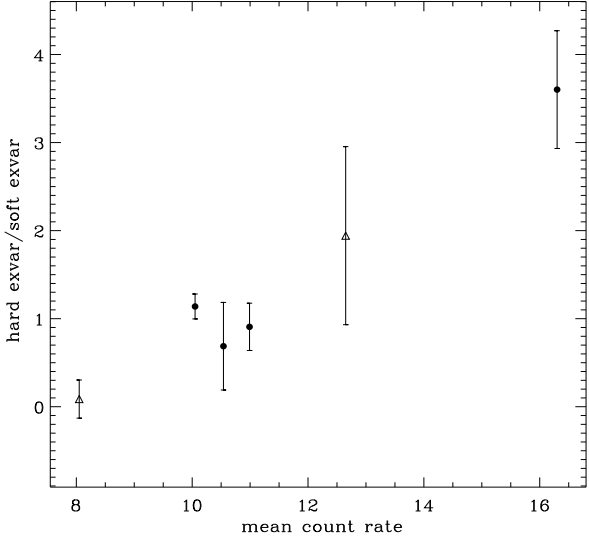,height=8cm,width=8cm,%
bbllx=40pt,bblly=70pt,bburx=550pt,bbury=600pt,clip=}
\caption{Ratio of excess variances versus total mean count rate. Each filled circle
is the average over the individual observation. The triangles represent
the average values over the two exposures during orbit 30. 
\label{figure:XMM57_f3}}
\end{figure}

Another interesting correlation was found between the variability in the
hard band and the total count rate . The variability was studied
using the excess variance (Nandra \cite{nand}), which is obtained by computing 
the variance of the overall light curve, subtracting the variance due to
measurement error and dividing by the mean squared. Care must be taken
in interpreting  the excess variance of different observations as it
is related to the length and the sampling of the time series (e.g. Leighly 
\cite{leig1}).
However this drawback is circumvented in our analysis by considering the
ratio between the excess variance in the hard band and that in the soft range
during each individual observation.
Fig.~\ref{figure:XMM57_f3} shows a clear correlation between the hard to soft 
excess variance ratio
and the mean count rate in the 0.2-10 band: the contribution of the hard band 
to the variability increases and becomes dominant as the mean count rate 
increases.
From a direct inspection of the individual values of the excess variance in
the soft and hard band, this behaviour turns out to be due more to the
actual increase of the hard variability than to a depletion of the soft
variability level. 

Based on ROSAT and ASCA data Fiore \cite{fior}  and Leighly (\cite{leig2}) reported
the presence of correlations between variability and X-ray luminosity 
and the steepness of the X-ray spectrum. A first
analysis seems to indicate that the XMM data of \object{PKS~0558-504}
are consistent with the previous results, however a detailed study of such
correlations is beyond the scope of this paper. A spectral study
of the XMM observations of \object{PKS~0558-504} can be found in O'Brien et al. (2001).

\begin{figure}
\psfig{figure=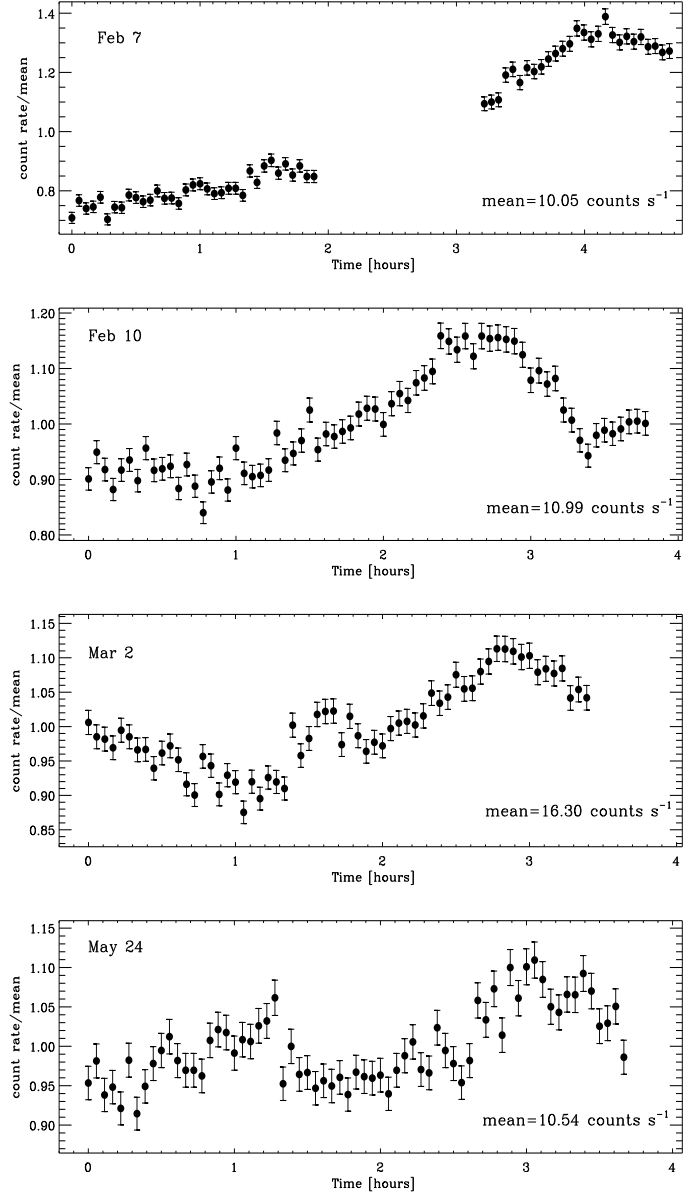,height=16cm,width=9cm,%
bbllx=15pt,bblly=30pt,bburx=500pt,bbury=812pt,clip=}
\caption{Individual EPIC PN 0.2-10 keV light curves of \object{PKS~0558--504} for 
orbit 30, 32, 42 and 84 with time binning of 200 s. 
\label{figure:XMM57_f4}}
\end{figure}

\section{Short term variability}
The \object{PKS~0558--504} EPIC PN light curves for each individual 
observations, binned into 200 s intervals, are shown in Fig.~\ref{figure:XMM57_f4}.
Strong and persistent X-ray flux variability is observed during each
observation. The most prominent increase in the count rate is observed
during orbit 30 (February 7), when the 0.2-10 keV flux increased by a factor 2 
in less than 4 hours. On the other hand the most extreme temporal variation in 
the count rate 
occurred on February 10, when the count rate decayed by $18\%$ in 1500 seconds.
This value is in agreement with that measured by the EPIC MOS during orbit 84
(O'Brien \cite{obri}), which led to the calculation of a lower bound for the 
radiative efficiency 
which exceeds the theoretical maximum for a non rotating black hole, under the
assumption that the photon diffusion through a spherical mass of accreting
matter is dominated by Thomson scattering.

A visual examination of the individual light curves indicates that
all of them show a similar long term variability pattern with
a common rise time of about 2 hours
and, in some cases, small amplitude flares superimposed.
The best method to quantify 
time variability without the problems encountered in the traditional Fourier 
analysis technique in case of unevenly sampled data is a  structure function
analysis (e.g. Simonetti \cite{simo}, Hughes \cite{hug}). The 
first-order structure function is the mean deviation for data points separated 
by a time lag $\tau$, $SF(\tau)=\langle [F(t)-F(t+\tau)]^2\rangle$. One of the
most useful features of the structure function is its ability to discern the
range of time scales that contribute to the variations in the data set: the
characteristic time scales of the variability are identified by the maxima 
and slope changes in the $\tau-SF$ plane. For a stationary random process the
structure function reaches a plateau state for lags longer than the longest
correlation time scale. If a light curve contains cycles of period $P$, the
$SF$ will rise to a maximum at $\tau=P/2$ (Smith \cite{smit}).

\begin{table}[t] 
\caption{Variability properties of the individual light curves.}
\begin{center}
\begin{tabular}{lllll}
\hline
\hline
\noalign{\smallskip}
Orbit & Count rate$^{a}$ & $\sigma_{\rm rms}^{b}$ &
$\sigma_{\rm rms}^{\rm soft}$ & $\sigma_{\rm rms}^{\rm hard}$ \\
\noalign{\smallskip} 
\hline
\hline
\noalign{\smallskip}
30a & $8.05\pm0.51$ & $3.3\pm1.2$ & $4.6\pm1.8$ & $0.4\pm2.0$\\
\noalign{\smallskip}
30b &  $12.65\pm0.78$ & $3.3\pm0.9$ & $2.7\pm1.0$ & $5.2\pm5.0$\\
\noalign{\smallskip}      
30tot &  $10.05\pm2.39$ & $55.0\pm3.0$ & $53.5\pm3.7$ & $61.0\pm14.6$\\
\noalign{\smallskip}      
\hline
\noalign{\smallskip}
32 & $10.99\pm0.95$ & $7.0\pm0.9$ & $7.9\pm1.3$ & $7.1\pm4.0$\\
\noalign{\smallskip}             
\hline
\noalign{\smallskip}
42 & $16.30\pm1.02$ & $3.6\pm0.5$ & $2.7\pm0.5$ & $9.7\pm3.0$\\
\noalign{\smallskip}             
\hline
\noalign{\smallskip}
84 & $10.54\pm0.51$ & $1.8\pm0.5$ & $2.0\pm0.7$ & $1.4\pm1.9$\\
\noalign{\smallskip}
\hline
\end{tabular}
\end{center}
$^{a}$ The errors on the count rate represent the dispersion around the mean.\\
$^{b}$ Excess variances are in $10^{-3}$ units.\\
\end{table}

\begin{figure}
\psfig{figure=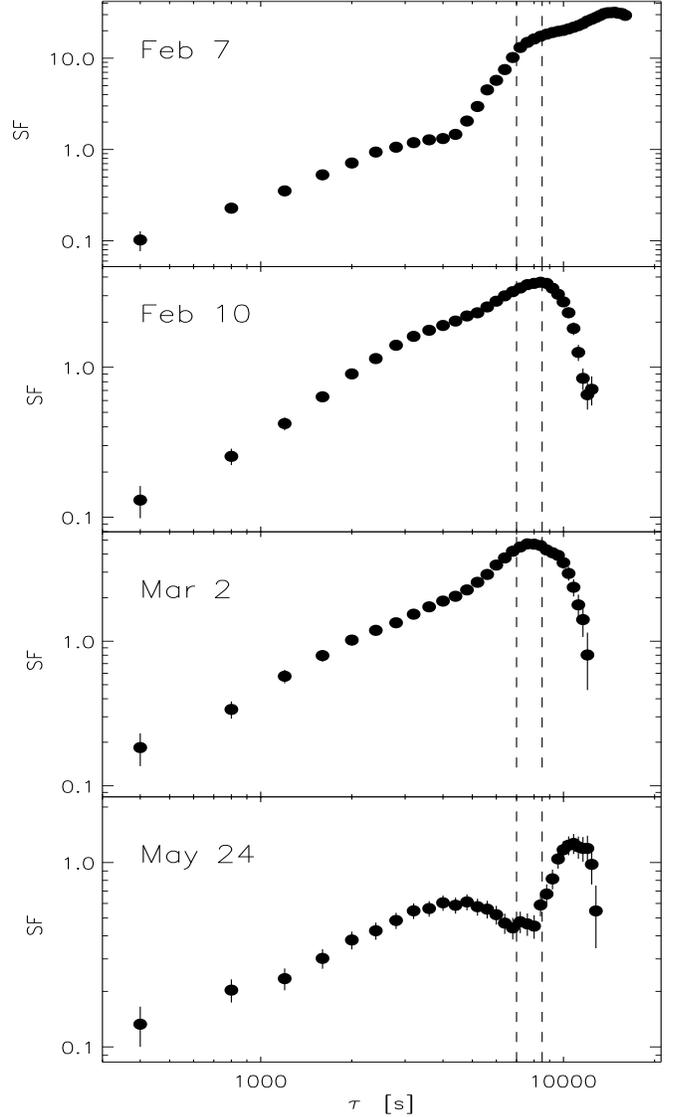,height=15cm,width=8.7cm,%
bbllx=130pt,bblly=85pt,bburx=430pt,bbury=725pt,clip=}
\caption{Structure functions of orbits 30, 32, 42 and 84. Time lags are 
in seconds. The dashed lines have been drawn to mark the interesting 
time lags interval. Most of the error bars are smaller than the symbol size.
 \label{figure:XMM57_f5}}
\end{figure}

Fig.~\ref{figure:XMM57_f5} shows that all the observations have either a
relative or absolute maximum around 2 hours. This means that a common typical 
time scale (which probably reflects the similar rise time in the light curves)
characterizes all observations. In addition, during orbits
30 and 84 a further characteristic time scale around 3 hours seems to be
present. A first confirmation for the presence of quasi-periodicity in the temporal
behaviour of \object{PKS~0558--504} comes from a periodogram analysis, which yielded
a strong signal at 2.3 hr. However, further longer observations are necessary
for a firm confirmation.

\begin{figure}
\psfig{figure=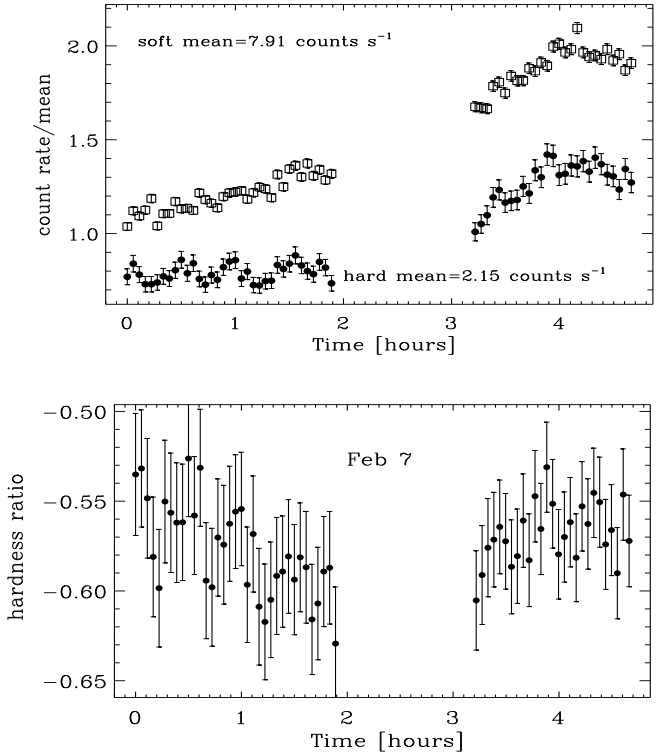,height=10cm,width=8.7cm,%
bbllx=15pt,bblly=70pt,bburx=490pt,bbury=715pt,clip=}
\caption{Top panel: soft (0.2-1 keV) and hard (1-10 keV) light curves during 
orbit 30 ; data are binned in 200 s intervals. The soft light curve divided 
by the mean (empty squares) has been multiplied by 1.5, to avoid
overlapping with the hard one (filled circles). Bottom: hardness ratio versus
time with 200 s binning.
\label{figure:XMM57_f6}}
\end{figure}

\begin{figure}
\psfig{figure=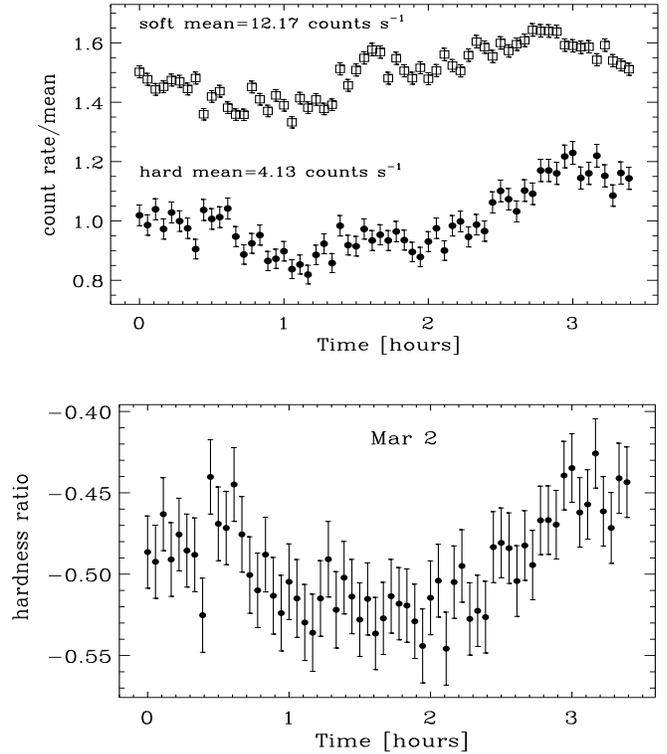,height=10cm,width=8.7cm,%
bbllx=15pt,bblly=70pt,bburx=490pt,bbury=715pt,clip=}
\caption{Top panel: soft (0.2-1 keV) and hard (1-10 keV) light curves during 
orbit 42 ; data are binned in 200 s intervals. The soft light curve divided 
by the mean (empty squares) has been multiplied by 1.5, to avoid
overlapping with the hard one (filled circles). Bottom: hardness ratio versus
time with 200 s binning.
\label{figure:XMM57_f7}}
\end{figure}

In order to seek evidence for spectral variability and for its origin
on short time scales we have split each of the broad band light curves into a 
hard and a soft light curve and plotted the hardness ratios versus the
time. While during orbits 32 and 84 (when the flux was at intermediate values)
no significant spectral variability
was found, orbits 30 and 42 present an interesting and somewhat puzzling
spectral and temporal behaviour shown in  Fig.~\ref{figure:XMM57_f6} and
Fig.~\ref{figure:XMM57_f7}, respectively. The mean count rates and the excess
variances in the soft, hard and broad bands are summarized in Table 2.

During the first exposure of orbit 30 the variability, quantified by the
excess variance $\sigma^2_{\rm rms}$ is strongly dominated by the soft
photons ($4.6\times10^{-2}~{\rm vs.}~4.0\times10^{-3}$, for the soft and
the hard band respectively), with
the soft flux steadily increasing and the hard fluctuating around a constant 
value. As a consequence, while the total flux increases
the hardness ratio decreases in time, with a slope of $(-2.9\pm 0.7)\times
10^{-2}$/hr obtained from a linear least square fit.
A totally different trend is observed during the second exposure, starting just
one hour later. In this case the variability is dominated by the hard flux
($\sigma^2_{\rm rms}:5.2\times10^{-2}~{\rm and}~2.7\times10^{-2}$, for the
hard and soft band, respectively), which increases faster than the soft emission
and seems to peak before it. The hardness ratio is steadily increasing with
the total flux and the time (slope: $(1.2\pm 0.4)\times10^{-2}$/hr).
A similar spectral behaviour is observed during orbit 42: the variability
is strongly dominated by the hard flux ($\sigma^2_{\rm rms}:9.7\times10^{-2}~
{\rm vs.}~2.7\times10^{-2}$) and the spectrum becomes harder when the
total count rate increases (slope: $(1.3\pm 0.3)\times10^{-2}$/hr).

On the basis of the five exposures available (two for orbit 30 and
one each for orbits 32, 42 and 84), the spectral variability on short
time scales of \object{PKS~0558--504}
can be summarized in the following way: when the average
broad band count rate is at a low level (the mean value is around
$8{~\rm cts~s^{-1}}$), the variability is dominated by the soft photons
and the hardness ratio decreases as the total flux raises. Conversely,
when the average count rate is above a certain level 
($12.6{~\rm cts~s^{-1}}$ and
$16.3{~\rm cts~s^{-1}}$ for the second exposure of orbit 30 and orbit 42, 
respectively), the variability is dominated by the hard band and the spectrum
becomes harder when the count rate increases. At intermediate values of
count rate (the mean is around $10{~\rm cts~s^{-1}}$ both for
orbit 32 and 84), the variability in the two bands is very similar
and no significant spectral variability is detected.

\section{Summary and conclusions}
We have presented the results of the EPIC PN observations of the
NLS1 \object{PKS~0558--504} taken during the Cal/PV phase. We found:
\begin{itemize}

\item During the XMM pointings \object{PKS~0558--504} showed a moderate but
persistent variability both on long (months) and short (minutes and hours) 
time scales.

\item Strong correlations between hardness ratio and total flux have
been found. When the mean flux is above a certain threshold,  
the spectrum becomes harder as the count rate increases. When the
flux is below a certain level, the spectral behaviour is in the opposite sense. 
At intermediate count rates no significant
spectral variability is observed.

\item A further strong correlation between the broad band flux and
the variability (expressed in terms of excess variance) in the hard energy 
band has been discovered: the contribution to the total variability of the 
hard band increases and becomes dominant as the mean count rate raises.

\item In all the observations the variability is characterized by smooth 
similar modulations, with typical time scale $\sim 2.2$ hr, 
determined with the structure function analysis.

\item The most extreme temporal variation of the count rate
implies a radiative efficiency
slightly larger than the theoretical maximum for accretion onto a 
Schwarzschild black hole. It is important
to point out that this is a strictly lower limit, since during our analysis only
single events (which represent $\sim 70\%$ of the total X-ray photons collected)
have been used. 
\end{itemize}

A detailed model of \object{PKS~0558--504} which accounts for all timing and 
spectral features discovered is beyond the scope of this paper. 
Therefore, we limit ourselves to some general considerations. 
The most intriguing result is the tentative detection of a common 
characteristic time scale of $\sim $ 2 hours. However a confirmation of this 
characteristic time scale has to await for longer observations, 
to sample a higher number of cycles. A more detailed analysis to assess its 
significance and robustness (e.g. as in Boller \cite{boll2}) is deferred to a
follow-up publication. This time scale
might be associated to periodic phenomena occurring in
the accretion disk, whose contribution seems to be dominant also
on the basis of the spectral analysis of EPIC MOS data (O'Brien \cite{obri}).
Two possible explanations which need to be
quantitatively tested  are: {\it i}) the presence of a hot spot
orbiting the black hole and {\it ii}) the precession of the inner part of the 
accretion disk caused by misalignment of black hole and disk rotational 
axes. An alternative 
tempting hypothesis involves the presence of a jet: as \object{PKS~0558--504} is a 
radio-loud object, beamed emission from a jet could partly contribute to the 
brightness and variability in X-rays. Radio-loud quasars are known to have
flatter X-ray spectral indices than radio-quiet ones, and they are also known
to be brighter X-ray sources (e.g. Yuan \cite{yuan}). A widely accepted 
explanation for this fact is that the X-ray emission includes an additional
harder component associated with the radio jet. In this scenario a naive
interpretation of the temporal and the spectral variability associated
with changes of the broad band flux, involves the presence of 
either a precessing jet or moving knots
with helical trajectories in a relativistic magnetized jet (Camenzind \&
Krockenberger \cite{came}). At this point it is worth noticing that also in
one other of the few radio-loud NLS1, RX J0134.2--4258, the spectral variability
(characterized, as in our case, by a hardening of the spectrum during the flux 
increase) was tentatively interpreted as  due to the jet emission (Grupe 
\cite{grupe}).

The exciting results from the temporal analysis of a few short exposures
demonstrate the extraordinary capabilities of XMM, in particular in light of the
long observations possible in virtue of the highly eccentric
orbit.

\begin{acknowledgements} 
The XMM - Newton project is supported by the Bundesministerium f\"ur
    Bildung und Forschung / Deutsches Zentrum f\"ur Luft- und Raumfahrt 
    (BMBF/DLR), the Max-Planck Society and the Heidenhain-Stiftung.
EPIC was developed by the EPIC Consortium led by the Principal Investigator,
Dr. M.J.L. Turner. The consortium comprises the following Institutes:
University of Leicester, University of Birmingham, (UK); CEA/Saclay,
IAS Orsay, CESR Toulouse, (France); IAAP Tuebingen, MPE Garching, (Germany);
IFC Milan, ITESRE Bologna, IAUP Palermo, (Italy). EPIC is funded by:
PPARC, CEA, CNES, DLR and ASI.    
MG  acknowledges support from
the European Commission under contract number ERBFMRX-CT98-0195 
(TMR network ``Accretion onto black holes, compact stars and protostars").

\end{acknowledgements}

\end{document}